\documentclass[12pt]{article}
\usepackage{amsfonts}
\usepackage{bm}
\usepackage{epsfig}
\usepackage[matrix,arrow,curve]{xy}
\usepackage{amsmath,amssymb}
\usepackage[utf8x]{inputenc}
\usepackage[english]{babel} 
\usepackage{graphicx}
\usepackage{mathenv}

\begin{document}

\pagestyle{plain}

\makeatletter
\@addtoreset{equation}{section}
\makeatother
\renewcommand{\theequation}{\thesection.\arabic{equation}}
\pagestyle{empty}
\begin{center}
\LARGE{Coleman de Luccia geometry reconsidered and ADS/CFT\\[10mm]}
\large{ José  A. Rosabal \\[6mm]}
\small{{\em Centro Atómico de Bariloche, Instituto Balseiro and CONICET} \\[-0.3em]
8400 S.C. de Bariloche, Argentina.\\[1cm]}
\small{\bf Abstract} \\[0.5cm]\end{center}
We reconsidered the Coleman de Luccia solution building an AdS$_4$ bubble expanding into a
false flat vacuum. In this construction when junction conditions are imposed we find an upper bound to
the radius of the AdS$_4$ and
a domain wall whose tension is a function of the minimum of the scalar potential. We prove that
this solution is exactly the solution found by Coleman and de Luccia, but in addition there is a
new condition that restricts the AdS$_4$ radius and a precise relation between the tension and the
minimum of the scalar potential. The applicability of the ADS/CFT correspondence is discussed.

{\small}

\newpage
\setcounter{page}{1}
\pagestyle{plain}
\renewcommand{\thefootnote}{\arabic{footnote}}
\setcounter{footnote}{0}


\section{Introduction}

In the study of eternal inflation \cite{inf1, inf2}, in String Theory, specifically in flux 
compactifications \cite{polc} and the study of the landscape 
problem \cite{kach, sus}, an important piece has been the Coleman de Luccia (CdL)
geometry \cite{cole1, cole2,cole3}. It has been recently proposed \cite{malda} that processes 
taking place in the CdL geometry can be dually described using the ADS/CFT correspondence 
\cite{malda1,klevpolik, witten}.

The study of this geometry is performed mostly in the thin wall approximation. This approximation
separates the space in two parts. One of the considered cases is when the interior part
is an anti de Sitter space and the exterior one is a flat 
space. These two parts are joined at the domain wall. 

Here we want to focus on the meaning of the thin wall approximation and some consequences that 
have not been taken into account  when this approximation is considered. The starting point of the 
study of the CdL geometry is the action
\begin{equation}
 16\pi S = \int_{M} (R-g^{\mu\nu}\partial_{\mu}\phi\partial_{\nu}\phi-V(\phi))\sqrt{-g}d^4x \label{ac}
\end{equation}
with no distinction \footnote{No distinction in the sense that $M$ is a one piece space and it 
is not an space formed by gluing several spaces.} over $M$, which is the manifold where the 
fields of the theory are defined. The key point is that as was pointed out above, when we assume 
the approximation of the thin wall the space breaks
into two parts, both with the same boundary at the domain wall. It is well known that when we study 
the theory of the gravitation on a manifold with boundaries, in order to get 
a well defined variation process the boundary terms have to be added in the gravitational action 
\cite{y, gh, hi}.

Due to this fact we claim that the boundary terms have to be added to  (\ref{ac}) in order 
to  be able to take the limit of the thin wall. If we study the CdL problem without the 
approximation these boundary terms, defined over a fictitious domain wall, do not contribute to 
the action (\ref{ac}).
This is because on  both faces of the fictitious domain wall the extrinsic curvature has the 
same value  and (\ref{ac}) does not change. 
On the other hand, if the study is performed on the mentioned approximation these boundary 
terms contribute to the action (\ref{ac}). This is because 
the manifold is metrically continuous but the domain wall becomes real and corresponds to the 
non-differential 
points of the metric and the normal derivative of the metric (extrinsic curvature) is different 
on both faces of the domain wall.

Our goal is to study the consequences of the inclusion of these terms in the CdL geometry. To this
end we organize the paper as follows. In Section $2$ we build the model starting with an action
equivalent to the one considered by Coleman and de Luccia, but with the boundary terms added. We 
pay special attention to the variational problem \cite{variation} of the action, which among others things
leads to  Israel's junction conditions \cite{israel, israel1}. We choose a different coordinate
 system to the used in \cite{cole3} and find a bound to the radius of the AdS$_4$ that has not been taken into account until 
now in the literature. In Section $3$ we solve the junction condition problem and find a precise 
relation on the 
tension of the domain wall and the minimum of the potential of the scalar field. In Section $4$
we prove that this solution is exactly the CdL one and we give the precise diffeomorphism 
that makes the identification possible. Finally in Section $5$ we present the conclusions together 
with an outlook and a discussion about the applicability of the ADS/CFT correspondence to 
describe processes taking place in the CdL geometry.

\section{The Bubble}

In this section we build the bubble solution. We choose to parameterize the whole manifold $M={\mathbb R}^4$
 by the usual spherical open patch $x^{\mu}=(r,\theta,\varphi,t)$.  
We divide the space in two regions $M_{-}$ and $M_{+}$. The region 
$M_{-}$ is defined by $\{x\in M:r<\rho(t)\}$ and $M_{+}$ by $\{x\in M:r>\rho(t)\}$. Here $\rho(t)$
is a function that will be determined by the junction conditions \cite{israel,israel1}. Of
 course, the existence of the distinction, $M_{-}$ and $M_{+}$ is conditioned by the existence
 of a solution of the junction conditions problem. 

We consider the action as
\begin{eqnarray}\label{action}
 16\pi S & =  & \int_{M_{-}} (R^{-}-{\mathcal L}(\phi_{-}))\sqrt{-g^{-}}d^4x+2\int_{\partial M_{-}}K^{-}\sqrt{-h^{-}}d^3\sigma+\\\nonumber
{} & {} & \int_{M_{+}} (R^{+}-{\mathcal L}(\phi_{+}))\sqrt{-g^{+}}d^4x-2\int_{\partial M_{+}}K^{+}\sqrt{-h^{+}}d^3\sigma+\\\nonumber
{} & {} & \int_{\partial M_{\infty}}2(K^{+}-K_{\infty})\sqrt{-h^{+}}d^3\sigma
\end{eqnarray}
where, ${\mathcal L}(\phi_{\mp})=g^{\mp\mu\nu}\partial_{\mu}\phi_{\mp}\partial_{\nu}\phi_{\mp}+V(\phi_{\mp})$,
$\partial M_{-}=\partial M_{+}\equiv M_0$ are the boundaries of $M_{-}$ and $M_{+}$ respectively
and correspond to the points $r=\rho(t)$.
The terms containing the scalar extrinsic 
curvature $K^{\mp}$  are
 included to get a well defined variation process. 
The boundary $\partial M_{\infty}$ at the infinity has an $S^3$ topology 
and $K_{\infty}$ is the extrinsic curvature of the boundary as embedded in flat space.

 We want to study only the vacuum solution such that the interior region
is an AdS space and the exterior is flat. In the thin wall approximation we choose the 
potential as ${\mathcal L(\phi_{-})}=V(\phi_{-})=-2\varLambda$ and
${\mathcal L(\phi_{+})}=V(\phi_{+})=0$. Here we do not perform the variation of the action (\ref{action}), 
see \cite{variation}, but to get $\delta S=0$ we have to impose the following constraints
\begin{eqnarray}\label{equ}
G^{\mp}_{\mu\nu}+\frac{1}{2}V(\phi_{\mp})g^{\mp}_{\mu\nu} & = & 0;\quad \forall x \in M_{\mp}\\\label{metpeg}
 h^{-}_{ab|_{M_0}} & = & h^{+}_{ab|_{M_0}}\\\label{constrain1}
\bigtriangledown_a\tau^{ab} & = & 0;\quad \forall \sigma \in M_{0} 
\end{eqnarray}
where $G^{\mp}_{\mu\nu}\equiv R^{\mp}_{\mu\nu}-\frac{1}{2}Rg^{\mp}_{\mu\nu}$ and
 $\tau^{ab}\equiv(K^{+ab}-K^{-ab})-(K^{+}-K^{-})h^{+ab}$. In the case $\tau^{ab}=0$ 
both spaces join smoothly. However, in the non-vanishing case we need
to include a new boundary term\footnote{We want to emphasize that 
when one computes $\tau^{ab}$ and finds that $\tau^{ab} \propto h^{ab}$, 
which is the case of a brane placed in the interface, the tension of this brane can not be chosen
freely. It has to be chosen in such a way that the energy momentum tensor $\frac{1}{2} T^{ab}$
 associated to the brane cancels out the $\tau^{ab}$ tensor.} such that
\begin{equation}
 \tau^{ab}+\frac{1}{2} T^{ab}=0;\quad \forall \sigma \in M_{0}\label{match}
\end{equation}
where $T^{ab}$ is the energy momentum tensor associated to the new boundary action constrained by
  $\bigtriangledown_aT^{ab}=0$. Moreover we have to impose the  Hamiltonian 
constraint on both faces of $M_0$ 
\begin{equation}
 (K^{+ab}K^{+}_{ab}-K^{+2})-(K^{-ab}K^{-}_{ab}-K^{-2})=2\Lambda;\quad \forall \sigma \in M_{0}\label{constrain2}
\end{equation}
which using (\ref{match}) can be rewritten as 

\begin{equation}
 \frac{1}{2}T^{ab}(K^{+}_{ab}+K^{-}_{ab})=-2\Lambda;\quad \forall \sigma \in M_{0}. \label{constrain3}
\end{equation}

Due to the apparent spherical symmetry of the studied problem we may naively choose the metrics  as $ds_ {\mp}^2=U^{\mp}(r)dr^2+r^2d\Omega^2_2-V^{\mp}(r)dt^2$ (in \cite{sg} this was the
 choice).  
However, here this choice is not the appropriate one. Invoking Birkhoff's theorem, the exterior solution will be
 the Schwarzschild one while the interior solution will be the usual static AdS.
Since the desired solutions are AdS-flat the appropriate choice is as follows

\begin{eqnarray}\label{met1}
 ds^2_{-} & = & f_{-}(r,t)(dr^2+r^2d\Omega^2_2-dt^2)\\
 ds^2_{+} & = & f_{+}(r,t)(dr^2+r^2d\Omega^2_2-dt^2)\label{met2}
\end{eqnarray}
where $d\Omega^2_2$ is the volume element of the unit $S^2$ sphere.

We can immediately see that the second metric is a solution of (\ref{equ}) on $M_{+}$ if 
$f_{+}(r,t)=1$. On the other hand it is very hard to solve the Einstein equation in the region $M_{-}$.
Nevertheless, without going into details of how the equation is solved on $M_{-}$, we see that the 
solution may be written as \footnote{See \cite{witten} for a discussion about the AdS space with this
 metric.}
\begin{equation}
 f_{-}(r,t)=\frac{4R^4_{AdS_4}}{(R^2_{AdS_4}+t^2-r^2)^2}\quad; \qquad R^2_{AdS_4}=\frac{3}{\Lambda}\label{funcion0}. 
\end{equation}
However we will not consider the last solution because it is impossible to glue consistently
 both spaces, which will be clarified in the next section. 
Instead and in order to glue both spaces we have to introduce a new scale $\epsilon$ and consider the solution as 
\begin{equation}
 f_{-}(r,t)=\frac{4\epsilon^4}{(\epsilon^4\frac{\Lambda}{3}+t^2-r^2)^2} . \label{funcion}
\end{equation}
It is a good exercise to see that the last expression is a solution of (\ref{equ}) on $M_{-}$.

Let us now explain some features of this $AdS$ space. In principle the introduced scale $\epsilon$ 
may be chosen arbitrarily. However, if we invoke String Theory we have to choose $\epsilon$
as the Planck length( $l_{p}$). In this patch the AdS$_4$ space can be seen as the region 
$r^2-t^2-l^4_{p}\frac{\Lambda}{3} \leq 0$ of the $\mathbb{R}^4$ space with metric $g^{-}$. The points
$r^2-t^2-l^4_{p}\frac{\Lambda}{3} = 0$ are harmless,
because these points correspond to the boundary of the AdS$_4$ space. In this patch the boundary is the 
hyperboloid
$r^2-t^2= l^4_{p}\frac{\Lambda}{3}$ and from the three dimensional point of view the boundary
may be seen as the expanding surface
$R(t)=\sqrt{t^2+l^4_{p}\frac{\Lambda}{3}}$. These features are used in the next section to discard
 one of the solutions 
that we find solving the junction condition problem.

\section{Junction Conditions}

In order to solve the junction condition problem we first look for a solution of 
 (\ref{metpeg}). It is easy to see that these two spaces can be glued only 
if $f_{-}(\rho(t),t)=1$  for all $t$. Let us focus on (\ref{funcion0}), solving
the last equation we have two solutions
\begin{eqnarray}\label{nonsol1}
\rho(t)^2-t^2 & = & -R^2_{AdS_4}\\ 
\rho(t)^2-t^2 & = & 3 R^2_{AdS_4}\label{nonsol2}.
\end{eqnarray}
The solution (\ref{nonsol1}) corresponds to the points of the $\mathbb{R}^4$ space placed 
in the non-causal region. On 
the other hand the solution (\ref{nonsol2}) does not fit into the AdS$_4$ space because 
the AdS$_4$ space with this metric ends in the hyperboloid $r^2-t^2=R^2_{AdS_4}$. For these reasons
we discarded (\ref{funcion0}). Of course we can fix this problem changing
the overall factor $1$ in front of (\ref{met2}). However, this is not an option because physically
we do not expect a scale in the flat space.

 Now we will solve $f_{-}(\rho(t),t)=1$
considering the metric on $M_{-}$ as (\ref{funcion}). The last equation has two solutions 
\footnote{The same solution was found in \cite{wittenbubble}.} well defined for all $t$
\begin{eqnarray}\label{solro1}
\rho(t) & = & \sqrt{t^2+\omega^2_{+}},\qquad \omega^2_{+}=l^2_{p}(l^2_{p}\frac{\Lambda}{3}+2)\\
\rho(t) & = & \sqrt{t^2+\omega^2_{-}},\qquad \omega^2_{-}=l^2_{p}(l^2_{p}\frac{\Lambda}{3}-2).\label{solro}
\end{eqnarray}
The solution (\ref{solro}) is well defined for all $t$ with the condition\footnote{We study only the
 case $(l^2_{p}\frac{\Lambda}{3}-2)>0$, when  $(l^2_{p}\frac{\Lambda}{3}-2)=0$, the surface 
expand at the speed of light and other treatment is required, see \cite{israel1}.}
 $(l^2_{p}\frac{\Lambda}{3}-2)\geq0$. Before continuing let us focus on 
(\ref{solro1}). This solution corresponds to the points of the hyperboloid $r^2-t^2= \omega^2_{+}$.
In the last section we saw that the AdS$_4$ space ends in the hyperboloid $r^2-t^2= \omega^2_{0}=l^4_{p}\frac{\Lambda}{3}$, so 
it is not difficult to see that (\ref{solro1}) does not fit in the AdS$_4$ space.
Discarding (\ref{solro1}), we will focus on (\ref{solro}). 

It is interesting that this solution needs an extra condition
to get a well defined one. The last condition is the most important result 
of this paper. To my knowledge it is a  bound that has not been taken into account until now
in the literature. 

Computing the scalar curvature associated to $g^{-}$ (using (\ref{funcion})) we find $R^{-}=-4\Lambda$. Since
the radius of curvature of the AdS$_4$ space may be written as $R^2_{AdS_4}=\frac{12}{|R^{-}|}=\frac{3}{\Lambda}$  
the bound may be rewritten as
\begin{equation}
 \frac{l^2_{p}}{R^2_{AdS_4}}>2.\label{bound}
\end{equation}
This is a very restrictive condition on the radius of the AdS$_4$ space. It goes against the 
possibility of using \cite{malda} the ADS/CFT correspondence to describe processes that take place in this
background.

The next step is to compute the tensors $K^{\mp ab}$ and the scalar extrinsic curvature 
$K^{\mp}=h^{ab}K^{\mp}_{ab}$. We will use the definition
\begin{equation}
 K_{ab}=(\bigtriangledown_{\mu} n_{\nu})e^{\mu}_ae^{\nu}_b
\end{equation}
where $n_{\mu}$ is the unit normal vector to the surface $M_0$, 
\begin{equation}
n_{\mu|_{M_0}}=\frac{1}{\sqrt{1-\dot{\rho}(t)^2}}(1,0,0,-\dot{\rho}(t)), 
\end{equation}
$\bigtriangledown_{\mu}$ is the covariant derivative and $e^{\mu}_a$ are the tangent vectors to $M_0$, 
\begin{eqnarray}
 e^{\mu}_{\theta} & = & (0,1,0,0)\\\nonumber
 e^{\mu}_{\varphi} & = & (0,0,1,0)\\\nonumber
 e^{\mu}_t & = & (\dot{\rho}(t),0,0,1).
\end{eqnarray}
After some algebra we find
\begin{equation}
 K^{-}_{ab}= (l^2_{p}\frac{\Lambda}{3}-1)K^{+}_{ab}=\frac{(l^2_{p}\frac{\Lambda}{3}-1)}{\omega}h_{ab}
\end{equation}
\begin{equation}
 K^{-}=(l^2_{p}\frac{\Lambda}{3}-1)K^{+}=\frac{l^2_{p}\Lambda-3}{\omega}\label{extrinsic}
\end{equation}
and
\begin{equation}
 \tau_{ab}=2\frac{\omega}{l^2_{p}}h_{ab}=2(\frac{1}{R^2_{AdS_4}}-\frac{2}{l^2_{p}})^{\frac{1}{2}}h_{ab}
\end{equation}
where 
\begin{equation}
 h^{-}_{ab}=h^{+}_{ab}\equiv h_{ab}=\textrm{diag}\big((t^2+\omega^2),(t^2+\omega^2)sin^2(\theta),-\frac{\omega^2}{t^2+\omega^2}\big).\label{metricin}
\end{equation}
with $\omega=\omega_{-}$ for simplicity.
At this point we see that a new 
term in the action (\ref{action}) has to be included\footnote{In the reference
\cite{mf} the same method was used.}, due to the form of $\tau_{ab}$. We conclude 
that an object with a positive tension
\begin{equation}
 T=(\frac{1}{R^2_{AdS_4}}-\frac{2}{l^2_{p}})^{\frac{1}{2}}\label{tension}
\end{equation}
is placed in $M_0$. Hence the new boundary action is
\begin{equation}
 S_B=2K_0\int_{M_0}\sqrt{-h}d^3\sigma\quad;\qquad K_0=-2T.\label{nambugoto}
\end{equation}

The last step is to check that (\ref{constrain3}) is satisfied. Computing the $T_{ab}$ tensor
\begin{equation}
 T_{ab}=-\frac{2}{\sqrt{-h}}\frac{\delta}{\delta h^{ab}}S_{B}=-4T h_{ab}
\end{equation}
and using (\ref{extrinsic}), this is straightforward to see that it is the case.
Rewriting the tension as $T=\frac{1}{l_{p}}(\frac{l^2_{p}}{R^2_{AdS_4}}-2)^{\frac{1}{2}}$ we can see
that the tension goes as $1/l_{p}$. So the choice in the last section $\epsilon=l_{p}$
is the  correct one since from String Theory the $2-$brane tension has a unit of $1/l_{p}$ 
\footnote{It is the correct unit considering the  physical constants $G=c=\hslash=1$.}.

Going further, we can ask the following two questions, i.e:
\begin{itemize}
\item What topology does the interface between $M_{-}$ and $M_{+}$ have?
\item What kind of space is this interface ?
\end{itemize}
The answers are simple, the 
interface in the three dimensional space can be seen as an expanding $S^2$ sphere 
with no restriction over the time coordinate. In others words, the variation range of the time is 
from infinity passed to infinity future. Thus the topology of $M_0$ is 
$S^2\times \mathbb{R} $. On the other hand, computing the scalar curvature associated
to (\ref{metricin}) we get  $R^{(3)}=6/\omega^2$. Hence we can identify $M_0$ as a three
 dimensional de Sitter space, see \cite{he} and references therein.

\section{Coleman de Luccia Solution}

In the last section we got a set of results that look like
the solution found in \cite{cole3}, written in other coordinates. To compare both solutions
we need to perform an diffeomorphism over the metric on $M_{-}$ and $M_{+}$ and check
that we can obtain metrics such as:
\begin{eqnarray}\label{mecol}
 ds^2_{-} & = & (\frac{3}{\Lambda})(d\xi^2_{-}+\sinh^2(\xi_{-})(\cosh^2(\eta)d\Omega^2_2-d\eta^2))\\\nonumber
ds^2_{+} & = & d\xi^2_{+}+\xi^2_{+}(\cosh^2(\eta)d\Omega^2_2-d\eta^2). 
\end{eqnarray}
Actually this is not difficult to do. Merely, over $M_{-}\cup M_{0}$ the transformation is given by
\begin{eqnarray}
 r & = & l^2_{p}\sqrt{\frac{\Lambda}{3}}\tanh(\frac{\xi_{-}}{2})\cosh(\eta)\\\nonumber
t & = & l^2_{p}\sqrt{\frac{\Lambda}{3}}\tanh(\frac{\xi_{-}}{2})\sinh(\eta)
\end{eqnarray}
while over $M_{+}\cup M_{0}$ we have
\begin{eqnarray}
 r & = & \xi_{+} \cosh(\eta)\\\nonumber
t & = & \xi_{+} \sinh(\eta).
\end{eqnarray}

However we must be careful with these transformations. When we perform different diffeomorphisms over
different regions we have to be sure that in the overlap region both diffeomorphisms are well
glued. From (\ref{mecol}) we see that both metrics
join continuously when 
\begin{equation}
 \sqrt{\frac{3}{\Lambda}}\sinh(\xi^{0}_{-})=\xi^{0}_{+}\label{pegado}
\end{equation}
 while both diffeomorphisms join continuously at $r=\rho(t)$  when $\xi^{0}_{+}=\omega$ and $cosh(\xi^{0}_{-})= (\omega ^2/l^2_{p})+1$, 
which is always satisfied due to the bound found. It is straightforward to check that the last two 
conditions are compatible with (\ref{pegado}). 
Computing the $\tau_{ab}$ tensor associated to the metrics (\ref{mecol}) we find 
 $\tau_{ab}=2 T^{'} h_{ab}$, where $T^{'}$ is the tension of the domain wall
\begin{equation}
 T^{'}=\sqrt{\frac{\Lambda}{3}}(\coth(\xi^0_{-})-\textrm{csch}(\xi^0_{-})).\label{ten}
\end{equation}
From (\ref{ten}) we can see that the tension of the domain 
wall, which may be seen as a 2-brane,  in this patch is a function of the position on the transverse 
coordinate. This is a negative fact from the CdL patch. This fact has been misunderstood by some authors who have 
considered that the brane tension can be chosen freely and then find the point $\xi^0_{-}$ where the brane is placed 
by solving the equations (\ref{ten}) and (\ref{constrain2}), in this patch. As we know from brane theory the 
brane tension can not be a function of the position on the transverse coordinate. Therefore the 
CdL patch is not the best candidate to perform the full study of the CdL geometry.
Using the result found in the other considered patch it is easy to see that 
\begin{equation}
T^{'}_{|_{\cosh(\xi^{0}_{-})= (\omega ^2/l^2_{p})+1}}=T.
\end{equation}

\section{Conclusions}

We have pointed out that the AdS$_4$ solution expanding into the flat space has a bound on
the radius of the AdS space. The studied configuration may evolve only if $l^2_{p}/R^2_{AdS_4}>2$.
We have obtained this bound due to the fact that we have looked for solutions to the junction conditions
well defined for all $t$. The non-well defined solutions have been discarded because they have
a temporal boundary and an initial singularity, breaking
down the distinction $M_{-}$ and $M_{+}$. 

We have taken seriously the variational problem of (\ref{action}). 
For this reason, instead of finding a bound to the tension of the domain wall, as
some authors have suggested \cite{bound, bound1}, we have found a precise relation between the tension (\ref{tension}) and the 
potential. This last relation is written in terms of the Planck length. This 
because the introduced scale $\epsilon$ has been considered as $\epsilon=l_{p}$. Although
we have invoked String Theory there is no proof that this statement is true.

 A promising outlook could be to embed the CdL geometry in a $10$-dimensional scenario and to use the 
full String Theory to compute the tension of the domain wall. The more important consequence of this bound, 
written in term of the Planck length or another scale, is that the bound tells us that  it is impossible to take the 
large $R_{AdS_4}$ limit and therefore it will be impossible to use the ADS/CFT correspondence to describe processes 
taking place in this background.

So far we have considered the thin wall approximation and we have used (\ref{action}) which is a different starting point
to the considered in \cite{cole3}. The key point is that this approximation separates the space in two regions, called 
in our case $M_{-}$ and $M_{+}$, and to get a well defined variation process inevitably we have to introduce the terms
that contain the scalar extrinsic curvature and the Nambu-Goto action (\ref{nambugoto}).
 Due to the last fact, the precise form of (\ref{action}), and the relation
(\ref{tension}) we can think about the whole setup studied here as an exact solution rather than an approximate one.

The proposal of \cite{malda} was taken with some criticism in reference \cite{suskin}. Here, 
using the bound (\ref{bound}) we have briefly discussed the applicability of the ADS/CFT 
correspondence to describe processes taking place in the CdL geometry and discarded this possibility. 
However, the most interesting thing is that although the AdS$_4$ radius has to be below 
the introduced scale, there is some kind of holography  in this setup because of the fact 
that the relation 
\begin{equation}
 T=(\frac{\Lambda}{3}-\frac{2}{\epsilon^2})^{\frac{1}{2}}\label{hol},
\end{equation} 
written in term of the scale $\epsilon$ (or the Planck length) 
is linking two parameters, $T$ and $\Lambda$, which are apparently  not linked.

Finally rewriting (\ref{hol}) as 
\begin{equation}
 T=\sqrt{\frac{\mid V(\phi_{-})\mid}{6}-\frac{2}{\epsilon^2}}<\sqrt{\frac{\mid V(\phi_{-})\mid}{6}}
\end{equation} 
 we can see that the tension $T$ satisfies the bound found in \cite{bound, bound1}, which is a necessary
 condition for the decay to proceed.

\section*{Acknowledgements}

We are grateful to Eduardo Andres, Gerardo Aldazabal, Horacio Casini, Daniel Harlow and Roberto 
Trinchero for useful discussions and comments. This work was supported by Consejo Nacional 
de Investigaciones Científicas y Técnicas, CONICET.

\bibliographystyle{my-h-elsevier}

\end{document}